\documentclass{mem}
\usepackage{natbib}\usepackage{txfonts}\usepackage{balance}
\usepackage{graphicx}
\usepackage[a4paper]{hyperref}
\idline{00}{1}
\begin{document}
\def\teff{$T\rm_{eff }$}
\def\kms{$\mathrm {km s}^{-1}$}

\title{Double Mode Cepheids with Amplitude Modulation}

\subtitle{}

\author{P.\thinspace Moskalik\inst{1} \and
        Z.\thinspace Ko{\l}aczkowski\inst{2} \and
        T.\thinspace Mizerski\inst{3}}

\offprints{P. Moskalik, \email{pam@camk.edu.pl}}

\institute{
Copernicus Astronomical Centre,
ul. Bartycka 18, 00--716 Warsaw, Poland
\and
Wroc{\l}aw University Observatory,
ul. Kopernika 11, 51--622 Wroc{\l}aw, Poland
\and
Warsaw University Observatory,
al. Ujazdowskie 4, 00--478 Warsaw, Poland
}

\authorrunning{Moskalik et al. }
\titlerunning{Modulated Beat Cepheids}

\abstract{Recent search for multiperiodicity in LMC Cepheids
(Moskalik, Ko{\l}aczkowski \& Mizerski 2004) has led to discovery
of periodic modulation of amplitudes and phases in many of the
first/second overtone (FO/SO) double mode pulsators. We discuss
observational characteristics and possible mechanisms responsible
for this behaviour.

\keywords{Stars: variables: Cepheids }
}

\maketitle{}

\section{FO/SO Cepheids of the LMC: Data and Analysis}

56 FO/SO double mode Cepheids have been identified in OGLE LMC
photometry (Soszy\'nski et~al. 2000). We have supplemented this
sample by 51 additional objects discovered by MACHO team (Alcock
et~al. 1999; 2003). The photometric data were analysed with a
standard prewhitening technique. First, we fitted the data with
double frequency Fourier sum representing pulsations in two radial
modes. The residuals of the fit were then searched for additional
periodicities. In the final analysis we used MACHO data (Allsman
\& Axelrod 2001), which offer considerably higher frequency
resolution than OGLE data.

\section{Results}

Resolved residual power close to the primary pulsation frequencies
was detected in 20 FO/SO double mode Cepheids (19\% of the sample).
These stars are listed in Table\thinspace 1.

\begin{table}
\caption{Modulated double mode Cepheids}
\vskip -0.3cm
\label{tab1}
\begin{center}
\begin{tabular}{lccc}
\hline
\noalign{\smallskip}
Star         & P$_1$  & P$_2$  & P$_{\rm mod}$ \\
             & [day]  & [day]  &  [day]   \\
\noalign{\smallskip}
\hline
\noalign{\smallskip}
SC1--44845   & 0.9510 & 0.7660 & ~~794.0  \\
SC1--285275  & 0.8566 & 0.6892 & ~~891.6  \\
SC1--335559  & 0.7498 & 0.6036 & ~~779.2  \\
SC2--55596   & 0.9325 & 0.7514 & ~~768.2  \\
SC6--142093  & 0.8963 & 0.7221 &  1101.6  \\
SC6--267410  & 0.8885 & 0.7168 & ~~856.9  \\
SC8--10158   & 0.6900 & 0.5557 &  1060.7  \\
SC11--233290 & 1.2175 & 0.9784 &  1006.2  \\
SC15--16385  & 0.9904 & 0.7957 &  1123.1  \\
SC20--112788 & 0.7377 & 0.5945 &  1379.2  \\
SC20--138333 & 0.8598 & 0.6922 & ~~795.0  \\
2.4909.67    & 1.0841 & 0.8700 &  1019.7  \\
13.5835.55   & 0.8987 & 0.7228 &  1074.9  \\
14.9585.48   & 0.9358 & 0.7528 &  1092.5  \\
17.2463.49   & 0.7629 & 0.6140 &  1069.9  \\
18.2239.43   & 1.3642 & 1.0933 & ~~706.8  \\
22.5230.61   & 0.6331 & 0.5101 & ~~804.3  \\
23.3184.74   & 0.8412 & 0.6778 &  1126.0  \\
23.2934.45   & 0.7344 & 0.5918 & ~~797.6  \\
80.7080.2618 & 0.7159 & 0.5780 & ~~920.3  \\
\noalign{\smallskip}
\hline
\end{tabular}
\end{center}
\end{table}

Stars of Table\thinspace 1 display very characteristic frequency
pattern. In most cases, we detected two secondary peaks on
opposite sides of each radial frequency. Together with the radial
frequencies they form two {\it equally spaced frequency triplets}
(see Fig.\thinspace 1). Both triplets have {\it the same frequency
separation} $\Delta{\rm f}$. Such a pattern can be interpreted as
a result of periodic modulation of both radial modes with a common
period ${\rm P}_{\rm mod} = 1/\Delta{\rm f}$. In Fig.\thinspace 2
we show this modulation for one of the stars. Both {\it amplitudes
and phases} of the modes are modulated. {\it Minimum amplitude of
one mode coincides with maximum amplitude of the other}. These
properties are common to all FO/SO double mode Cepheids listed in
Table\thinspace 1.

\begin{figure}[]
\resizebox*{\hsize}{!}{\includegraphics[clip=true]{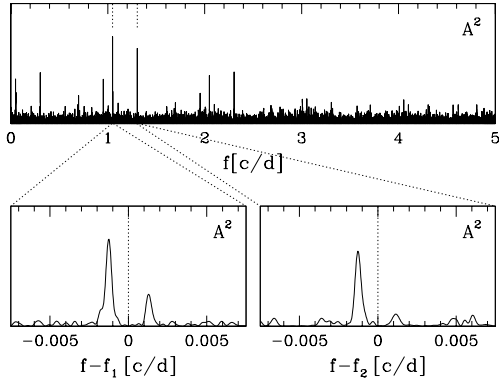}}
\vskip -1.5truecm
\caption{\footnotesize Power spectrum of LMC Cepheid SC1--44845 after
         prewhitening with two radial modes. Removed radial frequencies
         indicated by dashed lines. Lower panels display the fine structure.}
\label{fig1}
\end{figure}

\begin{figure}[]
\resizebox*{\hsize}{!}{\includegraphics[clip=true]{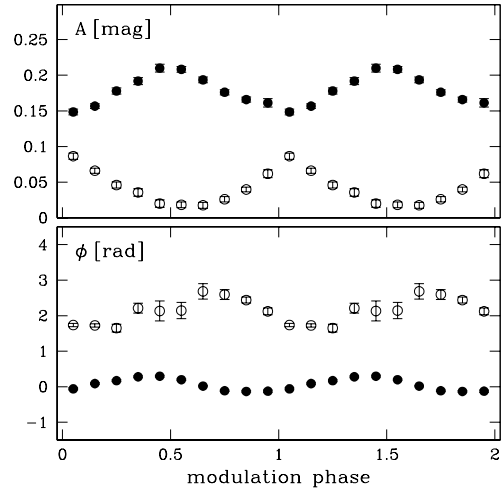}}
\caption{\footnotesize Periodic modulation of LMC double mode Cepheid
         SC1--285275. First and second overtones displayed with filed and open
         circles, respectively.}
\label{fig2}
\end{figure}

\section{What Causes the Modulation ?}

Two models have been proposed to explain Blazhko modulation in
RR~Lyr stars: the oblique magnetic pulsator model (Shibahashi
1995) and 1:1 resonance model (Nowakowski \& Dziembowski 2001).
Both models fail in case of modulated FO/SO double mode Cepheids,
being unable to explain why amplitudes of the two radial modes
vary in opposite phase (Moskalik et al. in preparation).

At this stage, the mechanism causing modulation in FO/SO Cepheids
remains unknown. However, common modulation period and the fact
that high amplitude of one mode always coincides with low
amplitude of the other, strongly suggest that energy transfer
between the two modes is involved. Thus, available evidence points
towards some form of mode coupling in which both radial modes take
part.

\vfill

\bibliographystyle{aa}

\begin{thebibliography}{}

\bibitem[1999]{Alc99} Alcock,~C. et~al. 1999, \apj,~511,~185

\bibitem[2003]{Alc03} Alcock,~C. et~al. 2003, \apj,~598,~597

\bibitem[2001]{All01} Allsman,~R.~A. \& Axelrod,~T.~S. 2001, astro-ph/0108444

\bibitem[2004]{Mos04} Moskalik,~P., Ko{\l}aczkowski,~Z. \& Mizerski,~T. 2004,
                      in Variable Stars in the Local Group, ed. D.~W.~Kurtz \&
                      K.~Pollard, IAU~Coll.~193, ASP~Conference
                      Series,~310,~498

\bibitem[2001]{Now01} Nowakowski,~R.~M. \& Dziembowski,~W.~A. 2001,
                      Acta Astron.,~51,~5

\bibitem[1995]{Shi95} Shibahashi,~H. 1995, in GONG'94: Helio- and
                      Asteroseismology from Earth and Space, ed. R.~K.~Ulrich,
                      E.~J.~Rhodes~Jr. \& W.~D\"appen, ASP~Conference
                      Series,~76,~618

\bibitem[2000]{Sos00} Soszy\'nski,~I., Udalski,~A., Szyma\'nski,~M., et~al. 2000,
                      Acta Astron.,~50,~451

\end{thebibliography}

\end{document}